\documentclass[iop]{emulateapj}
\usepackage{xcolor}
\usepackage{verbatim}
\usepackage{booktabs}
\usepackage{enumitem}
\usepackage{hyperref}
\usepackage[utf8]{inputenc}

\shorttitle{Tidal Imprints in the LMC Stellar Periphery}
\shortauthors{Besla et al.}
	
\begin{document}
	
\title{Low Surface Brightness Imaging of the Magellanic System: Imprints of Tidal Interactions between the Clouds in the Stellar Periphery} 

\author{Gurtina Besla\altaffilmark{1} , David Mart\'inez-Delgado\altaffilmark{2}, Roeland P. van der Marel\altaffilmark{3},
Yuri Beletsky\altaffilmark{4},  Mark Seibert\altaffilmark{5}, Edward F. Schlafly\altaffilmark{6,7}, Eva K. Grebel\altaffilmark{2},
Fabian Neyer\altaffilmark{8} }
\altaffiltext{1}{Steward Observatory, University of Arizona, 933 North Cherry Avenue, Tucson, AZ, 85721, USA}
\altaffiltext{2}{Astronomisches Rechen-Institut, Zentrum f{\"u}r Astronomie der Universit{\"a}t Heidelberg, M{\"o}nchhofstr. 12-14, D-69120 Heidelberg, Germany}
\altaffiltext{3}{Space Telescope Science Institute, 3700 San Martin Drive, Baltimore, MD, 21218, USA}
\altaffiltext{4}{Las Campanas Observatory, Carnegie Institution of Washington, Colina el Pino, 601 Casilla, La Serena, Chile}
\altaffiltext{5}{Carnegie Observatories, 813 Santa Barbara St, Pasadena, CA, 91101, USA}
\altaffiltext{6}{Lawrence Berkeley National Laboratory, One Cyclotron Road, Berkeley, CA 94720, USA}
\altaffiltext{7}{Hubble Fellow}
\altaffiltext{8}{ETH Zurich, Institute of Geodesy and Photogrammetry, 8093 Zurich, Switzerland}
\email{gbesla@email.arizona.edu}

\begin{abstract}

We present deep optical images of the Large and Small Magellanic Clouds 
(LMC and SMC) using a low cost telephoto lens 
 with a wide field of view to explore stellar substructure in the outskirts of the stellar disk 
of the LMC (r $<$ 10 degrees from the center).  These data have higher resolution than existing 
star count maps, and highlight the 
existence of stellar arcs and multiple spiral arms in the northern periphery, with no comparable 
counterparts in the South.  We compare these data to detailed simulations of the LMC disk
outskirts, following interactions with its low mass companion, the SMC. We consider 
interaction in isolation and with the inclusion of the Milky Way tidal field. The simulations are
used to assess the origin of the northern structures, including also the low density stellar 
arc recently identified in the DES data by \citet{Mac15} at $\sim$ 15 degrees. 
 We conclude that repeated close interactions with the SMC are primarily responsible for the 
 asymmetric stellar structures seen in the periphery of the LMC. The orientation and density of these
 arcs can be used to constrain the LMC's interaction history with and impact parameter of the SMC.  
 More generally, we find that such asymmetric structures should be ubiquitous about pairs of dwarfs
 and can persist for 1-2 Gyr  even after the secondary merges entirely with the primary. 
 As such, the lack of a companion around a Magellanic Irregular does not disprove the hypothesis 
 that their asymmetric structures are driven by dwarf-dwarf interactions. 
\end{abstract}

\section{Introduction}
\label{Sec:Intro}

The Large and Small Magellanic Clouds (LMC and SMC, respectively) are our
closest example of an interacting pair of dwarf galaxies. Evidence of this
interaction is clearly illustrated by the existence of the
gaseous Magellanic Bridge connecting the Clouds \citep[]{K57,H63}, and the 
tail of gas that trails them known as the Magellanic Stream \citep[]{M74,P03,N10}.
However, signatures of this interaction history are less clear in their
stellar components. 

On-going star formation in the LMC is stochastic, giving the 
dwarf galaxy an irregular appearance in the optical. For example, young, classical cepheids 
and binary stars trace out a single dominant spiral arm \citep{Moretti14}.   In contrast, near-IR 
surveys suggest that its older stellar population is smoothly distributed. 
The Two Micron All Sky Survey (2MASS) and Deep Near-Infrared Southern 
Sky Survey (DENIS) data for the LMC reveal a barred galaxy with a smooth old disk, 
extending to at least 9 kpc in radius \citep[]{vdM01}.  This is similarly 
revealed by the smooth distribution of RR Lyrae stars in the LMC disk \citep{Haschke2012}.

Asymmetries in the stellar disk do exist, however:  the disk appears to be warped \citep{vdMC01, OS02, N04} 
and the stellar bar of the LMC is geometrically off-center  
and warped relative to the disk plane \citep{Sub03, Lah05, K09}.
3D maps of the LMC created using Cepheids and RR Lyrae illustrate that the bar is 
in fact protruding from the disk \citep{Haschke2012,N04}.  This can be explained if the SMC
recently collided (impact parameter $< 5 $kpc) with the LMC disk \citep{B12}. 
This picture is supported by the discovery of 
low metallicity stars with distinct kinematics, consistent with SMC debris, in the LMC 
field \citep{CD12, O11,Graff2000,Kunkel97}. 

Models reproducing these asymmetric features in the LMC 
and the large scale gaseous structure of the Magellanic System 
require repeated tidal encounters between the LMC-SMC \citep[e.g.][]{B10,DB11}, 
Evidence of these repeated interactions should be more pronounced 
in the outer periphery of the stellar disk of the LMC, where the gravitational potential is shallower.  

Deep optical surveys of the stellar periphery of the Clouds are underway
with OGLE-IV \citep{U15} and the Dark Energy Camera on the Blanco 4 m Telescope, notably by the
Dark Energy Survey (DES)  and the Survey of the Magellanic Stellar History 
(SMASH). Such studies have revealed 
the presence of new dwarf galaxies, some of which may be companions 
\citep{DES1, K15, DES2, M15}, and stellar substructure that may trace the hierarchical 
assembly of the Clouds \citep[]{BK16}.  Such studies are complemented by ongoing 
surveys of the inner regions of the Clouds, such as the VISTA survey of the Magellanic Clouds 
\citep[VMC][]{Rubele15}. 

Intriguingly, the DES survey \citep[]{Balbinot15} and previous studies such as
the Outer Limit Survey \citep[]{S10} have also demonstrated that the LMC's
stellar disk extends much farther than the 9 kpc radius revealed
by earlier near-IR studies, potentially stretching as far as the Carina dwarf spheroidal galaxy, 
some 15-20 degrees away \citep[]{Mac15, Mc14, Mun06}.  
This implies that the disk of the LMC extends
at least 30 kpc in diameter. The tenuous periphery of such an extended structure 
is the ideal location to search for signatures of tidal perturbations from either the Milky Way 
or its binary companion, the SMC. Such evidence can be used to constrain the tidal radius,
and thereby the mass of the LMC, and is critical to constraining the interaction history with both the 
Milky Way and the SMC.

Recently, a potential stellar stream or arc was identified in the DES data 
in the northern outskirts of the LMC disk, $\sim$ 15 degrees from the LMC 
center \citep{Mac15}, stretching to the East. 
The authors suggest that this structure may be a result of Milky Way tides acting
 on the stellar outskirts of the LMC disk \citep[see also,][]{vdM01}.  
As their simulations illustrate \citep[Figure 12 in]{Mac15}, 
in such a scenario, both the northern and southern sections of the disk are affected:
there should be a matching structure in the South that stretches to the West.
The DES footprint does not extend to that section of the sky and the SMASH 
survey consists of a discrete set of non-contiguous pointings, which will be unlikely 
to pick out such faint structures.  In the near future, the DECam Magellanic Satellites Survey 
(PI: Bechtol) will extend the DES footprint to cover the southern LMC disk; this 
study may be able to shed more light on this question.

The DES results have not provided new information 
about the structure of the LMC's disk in the inner regions where the stellar density is higher.
The seminal optical observations of the LMC by 
\citet[]{deV72} (hereafter deVF72)
revealed that the LMC possesses pronounced 
multi-armed spiral structure in its northern stellar periphery. While the data quality is 
poor, these maps reveal that the outskirts 
of the LMC disk are not as smooth as the IR maps suggest. These spiral structures
continue to the outskirts of the disk and may play an important role in the origin of the 
structures observed by \citet[]{Mac15}.  Such details may be smoothed out in the DES maps.  
To address the nature of the northern arcs in the LMC disk, we thus require a different 
observing strategy than that employed to date. 

Ultra-deep, wide-field imaging 
using amateur telescopes can provide an alternative solution to 
map out substructure in the LMC disk out to large radii.
These small-size telescope data trace the substructures as diffuse light features, 
similar to the approach used in the stellar stream survey undertaken with 
similar facilities \citep{MD10}, with a typical surface brightness limit of 28.5 and 28 mag/arcsec$^2$
in $g$ and $r$. This is approximately three magnitudes deeper than the Sloan Digital 
Sky Survey II images. 
In particular, our team has mapped the stellar periphery of analogous systems
at much larger distances, such as the starbursting 
Magellanic Irregular galaxy, NGC 4449. Using a small
robotic telescope (0.5-m of aperture) and an exposure time of 18 hours in 
a luminance filter of the NGC 4449 system, 
\citet{MD12} revealed the existence of a faint stellar stream that may be the remains
of a disrupted low mass companion (half the mass of the SMC) orbiting 
about an LMC-mass dwarf galaxy located $\sim$4 Mpc away (see their figure 1).
Given the proximity of the Magellanic Clouds to the Milky Way ($\sim$50 kpc away), 
this observing strategy should also successfully reveal substructure in the 
stellar periphery of the Clouds, with the advantage of simultaneously image the 
entire Magellanic System -- including the entirety of both the LMC and SMC. 

Although these data are not as deep as the stellar density maps constructed from 
stellar tracers (blue horizontal branch or turnoff main sequence stars) selected in 
SMASH or DES color-magnitude diagrams, they serve an independent and
 complementary role of revealing substructure in the inner regions of the LMC's disk and how
such structure may propagate to the very outskirts.  

In what follows we present the first results of our panoramic imaging of the Magellanic
System (Section \ref{sec:Obs}) and compare these observations with simulations of the LMC disk 
that include repeated interactions with the SMC, with and without the tidal effects of the 
Milky Way (Section \ref{sec:Sims}) 
Our goal is to address the degree to which the northern arcs are seen in the south and 
whether interactions between the Clouds can reproduce the extent and degree of asymmetry 
observed without relying on the tidal influence of the Milky Way. 
Ultimately, we illustrate that such structures are to be expected around dwarf galaxy pairs 
in general (Section \ref{sec:Discuss}).

\section{A Deep Optical View of the Magellanic System}
\label{sec:Obs}

Our deep optical program makes use of small robotic telescopes, 
which provide long exposures at low cost. This observational strategy has been
used successfully to detect faint stellar streams about Milky Way analogs, as 
outlined in \citet[]{MD08,MD10,AM15} and about dwarf hosts \citep[]{MD12}. 

The deep optical imaging of the LMC and SMC fields presented here were obtained 
during a pilot optical program devoted to the search of stellar substructure around
some Milky Way dwarf satellites. Our observational strategy was designed to provide
unresolved images of these stellar systems, tracing their possible streams 
and other stellar substructures in their periphery by means of diffuse light 
detection. In addition, to avoid the foreground star and background galaxy 
contamination problems affecting the stellar density maps, we have adopted 
a technique that provides a fast (e.g. a single pointing) and cheap way to 
explore the periphery of these nearby galaxies, similar to those used in 
ultra-deep imaging of galaxies situated at some tens of Mpc away 
\citep[e.g.][]{Mihos05,MD10,Ab14,vDok14,Duc15,vDok15}.
Given that the distance of our targets is always less than some hundreds of kpc, 
we need to use very short-focal ratio instruments (i.e. f/3 or less)  
to avoid resolving the system into individual stars. In some cases,
this low resolution requirement means to use a high quality telephoto 
lens or apochromatic refractor telescopes, which, in the majority of the cases, 
is installed on a portable mount working in a very dark site. More details about this project 
and observational strategy will be available in an upcoming paper (Martinez-Delgado et al. in prep.). 

The imaging data of the LMC and SMC fields described in the following 
sections was obtained during two observing
 runs in August and September 2009 at ESO La Silla observatory. The imaging was done 
 using a portable setup consisting of a SBIG STL-11000M CCD camera and Canon 
 prime lenses: Canon EF 50mm f/1.4 USM and Canon EF 200mm f/2.8L II USM, which 
 yield the FOVs of 39 x 27 deg and 10 x 7 deg respectively. The corresponding pixel 
 scales are 37"/pixel and 9.27 "/pixel accordingly. Each image set consists of deep 
 multiple exposures obtained with a Baader Luminance filter (4000$\AA < \lambda < 7500\AA$). 
 For each pointing we also obtained a set of images with Baader red, green, and blue filters. 
 The individual exposure time in the Luminance filter was 300 sec (for Canon EF 200mm f/2.8L
  II USM lens) and 600 sec (for Canon EF 50mm f/1.4 USM lens). Standard data reduction 
procedures for bias subtraction and flat-fielding were carried out using the CCDRED 
package in IRAF.

\subsection{Panoramic View of the Magellanic System}

\begin{figure*}
\begin{center}
\mbox{\includegraphics[width=7in]{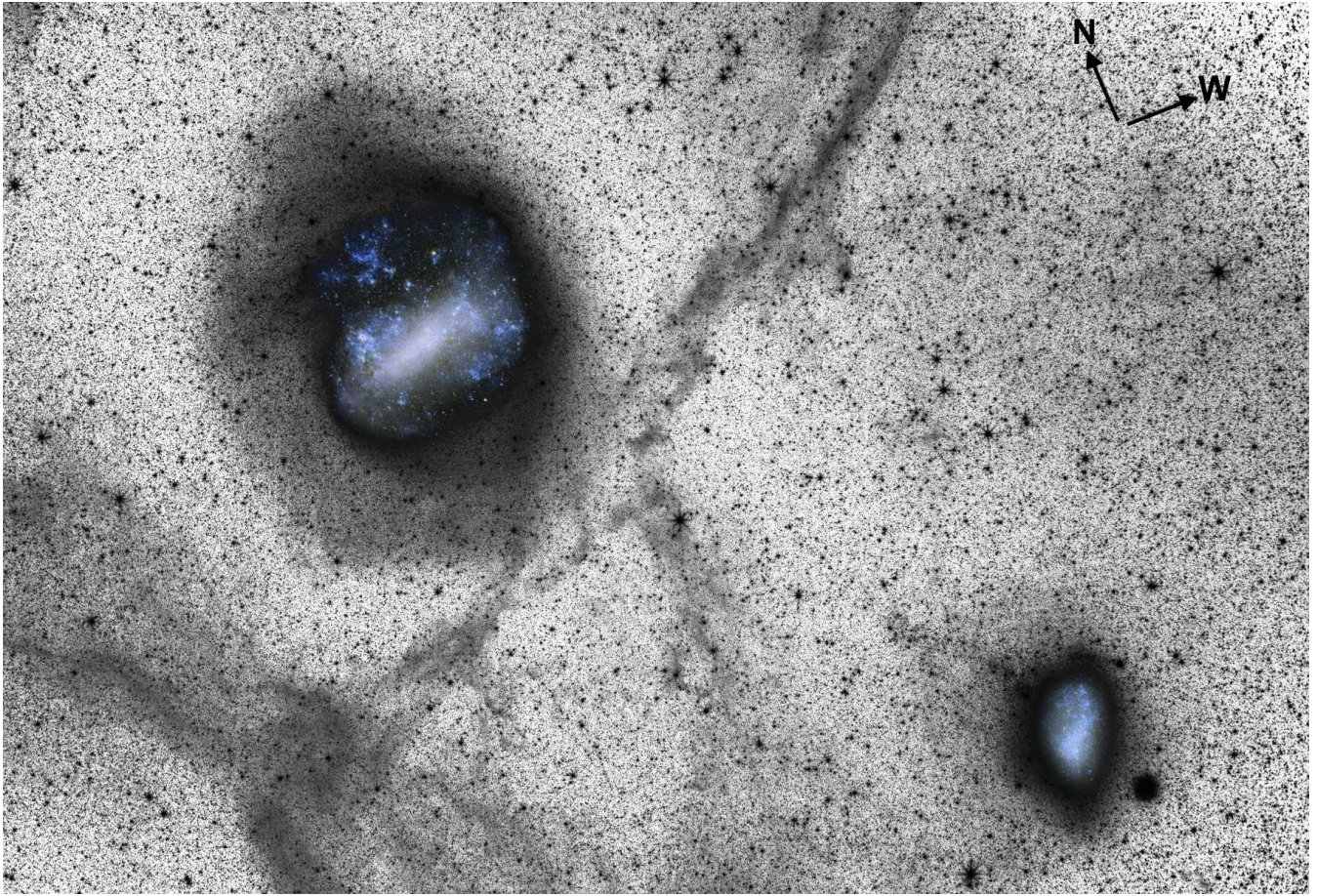}}
 \end{center}
 \caption{\label{fig:Panorama} Wide-field Luminance filter image of the Magellanic System
  (39 x 27 degrees). The LMC is located towards the top left and the SMC is to the bottom right. The 
  Milky Way globular cluster 47 Tuc is visible to the West of the SMC.
  A tail of stars from the SMC is visible stretching towards the LMC in the East.  The outskirts of the LMC disk 
  display pronounced asymmetries.  For illustrative purposes, a color inset of the inner regions of the 
  LMC and SMC, made from the color data obtained in our observing run (see Section~\ref{sec:Obs}), 
  is inserted as a reference and for comparison with previous studies. }
 \end{figure*}

% PANORAMA
Figure ~\ref{fig:Panorama} shows a zoomed section of the panoramic view of the Magellanic System made 
using the Canon EF 50 mm f/1.4 USM lens (Martinez-Delgao et al. in prep).  
To the West of the SMC is the globular cluster 47 Tuc.  
To the East of the SMC a tail of young stars extending $\sim$6 degrees towards the LMC is discernible. 
These stars are situated in the Magellanic Bridge and are most likely forming in situ, rather than being tidally stripped \citep{H07}. 

The LMC shows  asymmetric structure in its outskirts that is more pronounced
in the north (opposite to the direction of the SMC). These structures will be discussed 
in more detail in Section~\ref{subsec:LMC}.

\begin{figure*}
\begin{center}
\mbox{\includegraphics[width=3.5in, clip=true, trim= 6in 1in 1in 6in ]{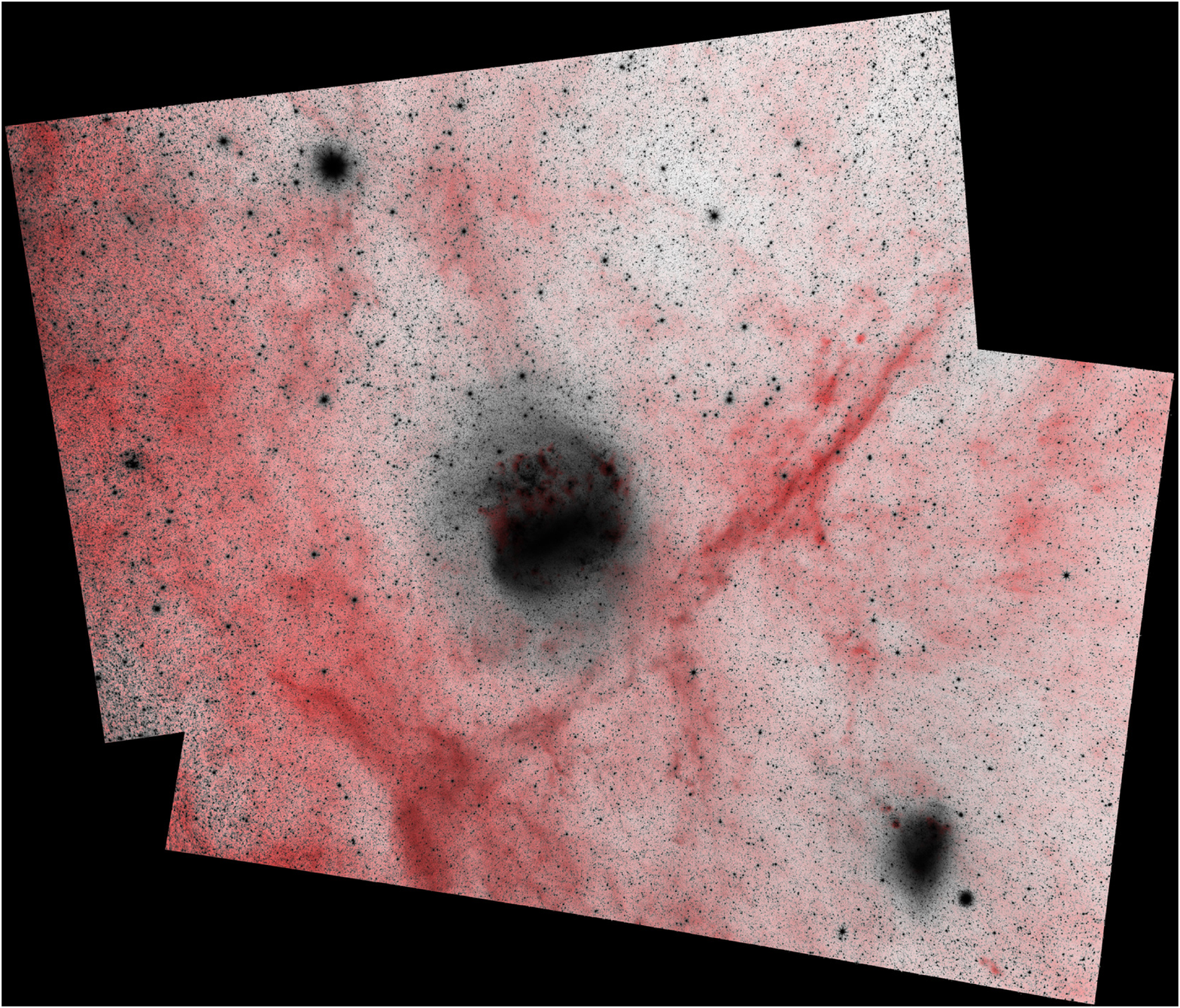}
\includegraphics[width=3.5in]{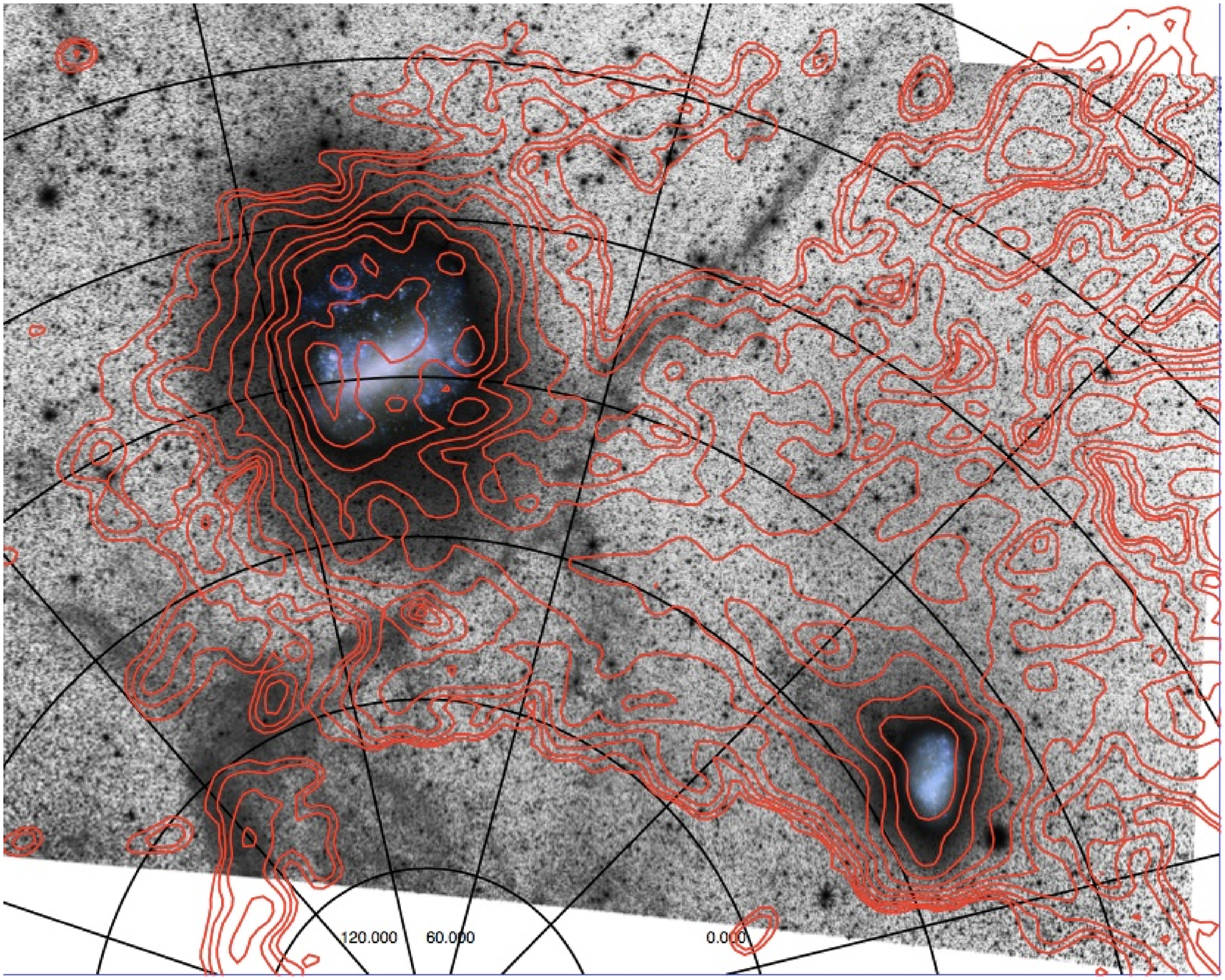}}
 \end{center}
 \caption{\label{fig:Cirrus} Left: Same as Figure~\ref{fig:Panorama} where Galactic cirrus, 
 identified based on far-IR observations by IRAS and DIRBE,
 is highlighted in red. There is very little cirrus in the outskirts of the LMC disk.
  Right:  Same with HI contours from \citep{P03} overplotted. Column densities
range from $10^{19}$ to $10^{21}$ atoms cm$^{-2}$.  There is a sharp drop in gas density
in the LMC outskirts towards the North East (top left), which was modeled in \citet{Salem15}
as an effect induced by ram pressure.   }
 \end{figure*}

% Cirrus 
Many of the structures visible in the image are Galactic cirrus, which are 
abundant at high galactic latitude in deep imaging with surface brightness 
limits fainter than 28 mag/arcsec$^2$. Some of these cirrus features 
(like the collimated filament between the LMC and the SMC) were previously
detected in the photographic plate by \citet{deV72}. 
We use the dust map of \citet[]{S98} to
disentangle surface-brightness features caused by dust from features
in the stellar density. These are highlighted in red in the left panel 
of Figure~\ref{fig:Cirrus}.   
The map is based on far-infrared observations
from IRAS and DIRBE, which are used to estimate the dust column
density from its brightness and temperature.  The map fails to
accurately track the dust column within the Large Magellanic Cloud,
where the temperature structure along each line of sight is more
complicated than the single, constant temperature assumed by \citet[]{S98}. 
However, the map still qualitatively
traces the dust throughout the region, allowing identification of
regions and features unaffected by the dust. From these maps it is 
clear that cirrus does not strongly affect the identification of
 structure in the LMC outskirts.   

The extended structures discussed in this paper are also seen in maps of stellar clusters, 
as illustrated by \citet[]{Bica08} and \citet{K90}, confirming that they are not artifacts 
of the image processing or Galactic cirrus.

% HI Overlay 
In the right panel of Figure~\ref{fig:Cirrus}, HI contours using data from \citet[]{P03} 
are plotted over the optical panorama in Figure~\ref{fig:Panorama}.  
The tail of young stars from 
the SMC is located in the highest column density regions of the bridge. 

There is a sharp drop off in the gas density towards the upper left (North East), 
which is the LMC's direction of motion. Our data illustrates that the stellar disk extends beyond 
this gas truncation radius.  This is likely the result of ram pressure stripping 
as the LMC moves through the circumgalactic medium \citep[CGM][]{Salem15}. Signatures of this process are likely 
visible in the truncation time of star formation in the outskirts of the LMC \citep{Meschin14}, particularly at the larger
distances probed by DES. 

There is a similar drop off in gas density to the South East of the SMC
(Pearson, Besla et al. submitted).  While our data does not suggest the main body of the SMC extends 
beyond this radius, \citet{N11} have illustrated that SMC stars extend to radii as large as 11 kpc, 
well beyond this gas drop off radius of a few degrees.

% l b r t

\subsection{ The Large Magellanic Cloud} 
\label{subsec:LMC}

\begin{figure*}
\begin{center}
\mbox{\includegraphics[width=7in, clip=true, trim=1in 6in 1in 2in ]{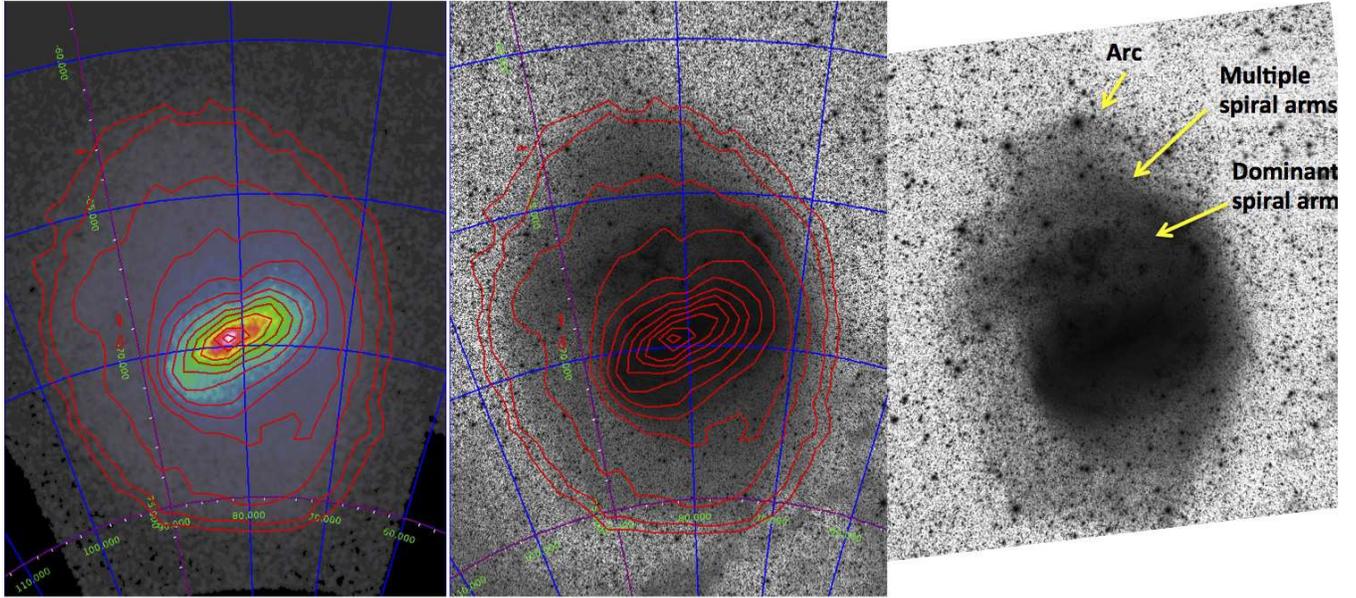}}
 \end{center}
 \caption{\label{fig:LMC} Left: 2MASS near-IR star count map from \citet{vdM01}. Countours extend
 to 8 degrees from the LMC center.  Center: Our LMC Luminance filter
 data with 2MASS contours over plotted. Right:  Luminance filter image, detailing the structure of the 
 LMC. Multiple ``embryonic" arms appear to the north of the dominant spiral arm. 
 The northernmost structure appears to be a pronounced stellar arc.  
 Comparable structures do not exist in the South.
 The stream found in the DES data by \citet{Mac15} is located above the top edge of these figures 
 (see Figure~\ref{fig:Mac}). } 

 \end{figure*}

In this section we focus on the LMC in more detail.
With that purpose, we use a higher resolution (9.3 arcsec/pix) image obtained
with the Canon EF 200 mm f/2.8 lens. Plotted in the right hand panel of 
Figure~\ref{fig:LMC} is the resulting $\sim$ 20 x 20 degree Luminance filter image of the 
LMC. The agreement between the observed morphological perturbations in the 
northern periphery of the LMC in this independent image and those in Figure~\ref{fig:Panorama}
confirms that they are not related to artifacts or background fluctuations (e.g. reflections
or flat fielding artifacts) from the data. 

Figure~\ref{fig:LMC} shows a comparison of the diffuse light structures visible in our deep 
image (right panel) with the 2MASS star count data \citep[left hand panel][]{vdM01}. 
In the center panel, contours from the 2MASS data are plotted over 
our Canon 50 image data. The 2MASS map involves a significant amount of smoothing, 
since the stellar density of red stars in the LMC
outskirts is low. Moreover, only old and intermediate-age (RGB and AGB) stars in the LMC
 were stacked to create the 2MASS maps.

Our data reaches a similar depth and extent as the 2MASS map. Based on the surface brightness
profile presented in \citet{S10}, this depth corresponds to $\sim$ 27 mag arc sec${^2}$.  The new 
direct image (right panel, Figure~\ref{fig:LMC}) reveals much more detail in the stellar disk than
the 2MASS data.   Multiple ``embryonic" arms are seen north of the dominant 
spiral arm, as noted by deVF72.   A pronounced stellar arc is visible $\sim$8 degrees North of the center 
of the LMC. These structures are barely visible in the deVF72 maps and are referred to
 as ``semi-detached outlying fragments of spiral arcs" (their region D).  The star $\beta$-Doradus
is seen in projection in the center of the northernmost arc.
 deVF72 suggest that this arc may be limited on the West and East sides by Galactic cirrus, 
 but Figure~\ref{fig:Cirrus} indicates this is unlikely.   
 
\begin{figure*}
\begin{center}
\mbox{\includegraphics[width=6in, clip=true, trim=1.2in 1.4in 1.4in 2.6in]{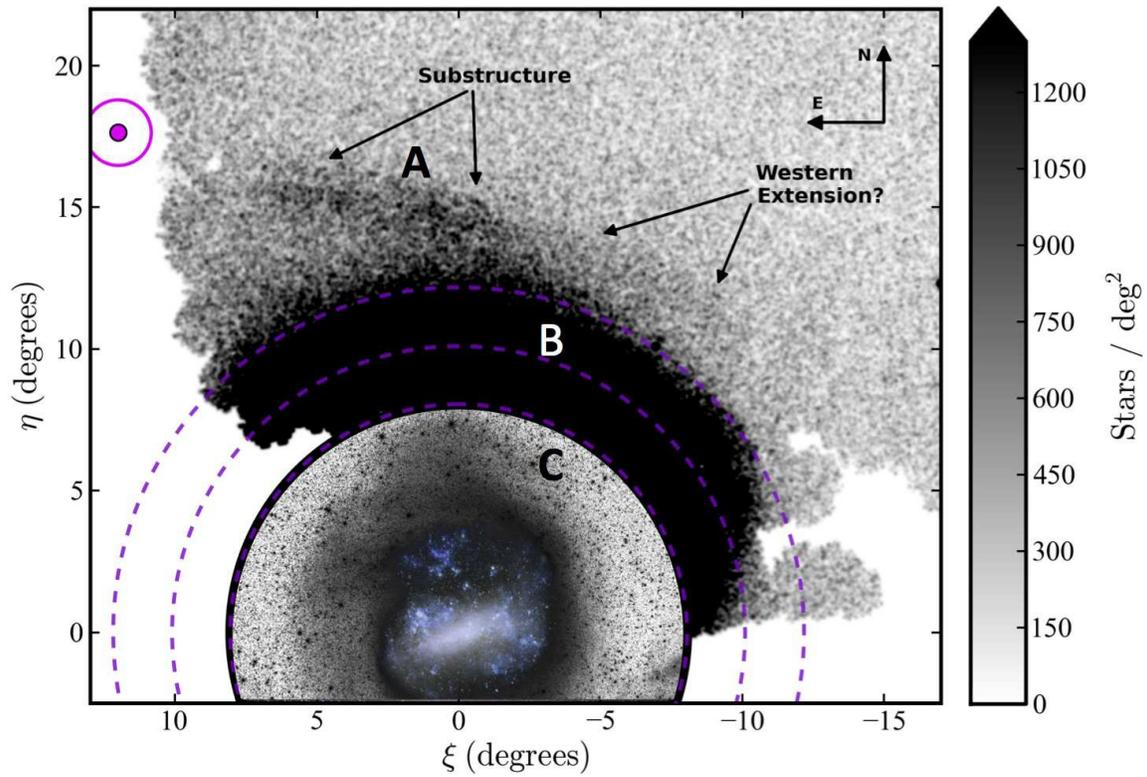}}
 \end{center}
 \caption{\label{fig:Mac} A modified version of Figure 1 from \citet{Mac15} showing the spatial 
 density of old main sequence turn-off stars in the LMC. the three purple dashed circles 
 indicate angular separations of 8$^{o}$, 10$^{o}$ and 12$^{o}$ from the center of the LMC.
 The stellar stream identified by \citet{Mac15} is marked as region A and extends towards the 
 Carina dwarf (magenta point).   Our data extends to $\sim$8 degrees. The substructure we have 
 identified in the LMC is marked as region C.  The intermediate region between our data and 
 the DES arc is marked as region B.} 

 \end{figure*}

 In Figure~\ref{fig:Mac} we have illustrate a modified version of Figure 2 in \citet{Mac15} with our 
color image of the LMC. 
Our data (marked as region C) do not extend as far as the DES star count data (which stretches to radii of 15 degrees), but
they illustrate that structures reminiscent of streams or arcs in the North begin at much smaller 
radii and do not have counterparts in the South. In particular, there are no streams that extend towards the 
East (right) in the South, suggesting that the Northern structures are unlikely to arise owing to a global 
tidal field from the Milky Way.  

We have marked three important regions in the DES data maps:  A) The northern-most arc or stream 
B) intermediate area between our data and the arc, where the stellar disk drops off sharply, 
C) regions that overlap with our data.  We now turn to simulations to interpret the origin of substructure 
in these regions.  

% l b r t

% l b r t

\section{Comparison with Simulations}
\label{sec:Sims}

\citet[][hereafter B12]{B12} postulate that asymmetric spiral structure 
in the LMC results from tidal interactions with the SMC, rather than interactions
with the Milky Way. They further argue that this scenario is generically
applicable to Magellanic Irregulars in the field. Thus, LMC analogs, regardless of environment, 
should also have (or once had) low mass companions. 

In the following, we explore the structural evolution of the LMC by examining simulations of the interaction 
and eventual merger of an isolated LMC and SMC binary pair of dwarfs (1:10 mass ratio
encounter; Section~\ref{subsec:LSsim}). 
We then explore how the resulting structures might be affected by the large scale tidal 
field of the Milky Way as the Clouds approach our Galaxy for the first time, by revisiting the simulations
presented in B12 (Section~\ref{subsec:LSMW}).  
Specifically, we focus on the structure of the stellar periphery of the 
LMC as it interacts with the SMC and compare these structures with the high resolution 
data presented in this paper.

\subsection{The LMC and SMC in Isolation}
\label{subsec:LSsim}
 
We begin by examining the impact of tidal perturbations on the LMC stellar periphery 
owing to repeated orbits of the 
SMC about an otherwise isolated LMC. We created simulations of this interaction using the 
smoothed particle hydrodynamics code, GADGET-3 \citep[]{S05}. These simulations were then 
used in B12 to model the evolution of the Clouds before they were captured by the Milky Way. 

Simulation parameters are outlined in detail in section 2 of B12. 
The Clouds are both modeled with exponential stellar and gas disks and 
live Hernquist dark matter halos.  
The stellar mass of the simulated LMC  is initially $2.5\times 10^{9}$M$_{\odot}$, its gas mass
is $1.1\times10^{9}$M$_{\odot}$ and its total dark matter mass is $1.8 \times 10^{11}$M$_{\odot}$, 
consistent with expectations from $\Lambda$CDM \citep{B10,BK11,M13,B15,G15,Pen16}.

The stellar mass of the simulated SMC is initiallly $2.6\times10^{8}$M$_{\odot}$, its gas mass is 
$7.9\times10^{8}$M$_{\odot}$ and total dark matter mass 
is $2.1\times 10^{10}$M$_{\odot}$. See Table 1, of B12 for full details of the numerical set up. 
This simulation accounts for star formation and thus the stellar mass increases over time, 
roughly approximating the current stellar mass of the Clouds today after $\sim$6-7 Gyr of evolution.  

\begin{figure*}
\begin{center}
\mbox{\includegraphics[width=7in, clip=true, trim=0.5in 7in 6.5in 0in ]{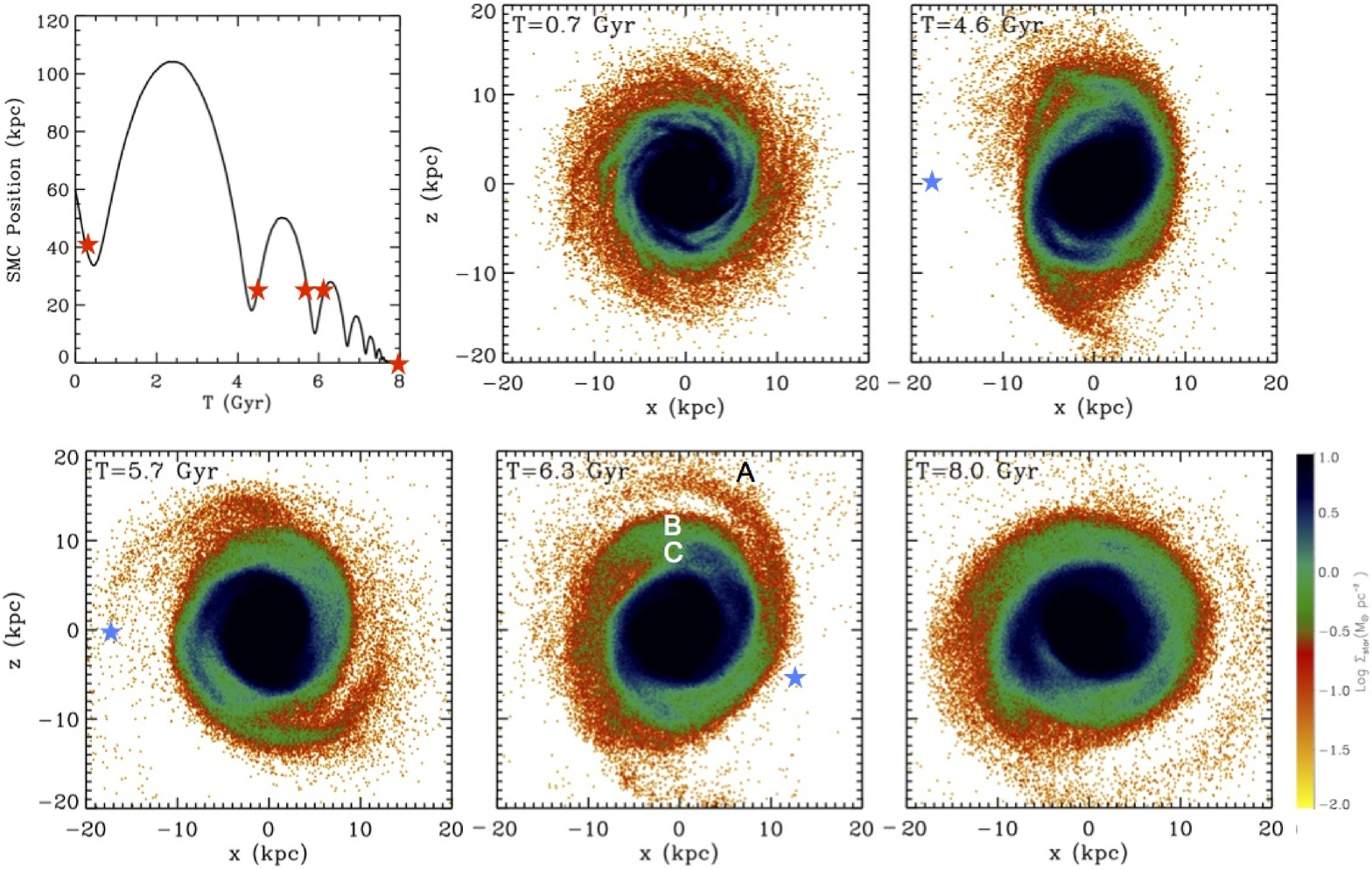}}
 \end{center}
 \caption{\label{fig:LSsim} The simulated interaction history of the Large and Small Magellanic Clouds in 
 isolation (i.e. without the Milky Way).  The top left panel illustrates the separation between the SMC and 
 LMC as a function of time. The Clouds were likely captured by the Milky Way after 5 or 6 Gyr of 
 evolution as an isolated binary pair. However, here the simulation is followed past that point until the system 
merges, as would have happened had the Clouds never been captured. Stellar density maps of the LMC disk 
are plotted at key moments in the interaction history as denoted by red stars in the top left panel. 
T=0.7 Gyr represents the initial state of the LMC disk as a symmetric exponential disk. At T=4.6, 5.6, and 6.3 Gyr
the SMC is roughly 25 kpc from the LMC, as it is today. SMC particles have been omitted from these maps, but 
the rough location of the SMC is marked by a blue star if it is visible in the field of view. 
 After 6.3 Gyr of evolution the SMC has just passed 
through the disk of the LMC ($b < 10$ kpc), inducing strong asymmetrical spiral structure and arcs (Region C) that exist as 
far as 15 kpc from the LMC center (Region A). This structure is similar to that seen in \citet{Mac15}, illustrating that low
density arcs can form at large radii from the LMC center without the influence of the Milky Way tidal field.   
After 8 Gyr of evolution the system has completely merged, yet asymmetric 
spiral structure still persist.  
    }
 \end{figure*}

The SMC is placed on an eccentric orbit about the LMC (eccentricity of 0.7), which slowly 
decays over time owing to dynamical friction. The separation between the galaxies is plotted as
a function of time in the top left panel of Figure~\ref{fig:LSsim}.
This figure is an extension of the orbit shown in Figure 2 of B12. In B12, the 
binary pair followed a trajectory that entered the Milky Way's virial radius after roughly 
5 or 6 Gyr. Here instead, we keep the LMC and SMC pair in isolation and allow the binary to evolve 
further in time until the SMC is eventually completely cannibalized by the LMC
(after 8 Gyr of evolution).  This represents the ultimate fate of the Magellanic Clouds if they had 
never been captured by the Milky Way and may mimic the evolution of pairs of dwarfs in the 
field.  The structures produced are therefore independent of the 
Milky Way's tidal field. 

The color panels in Figure~\ref{fig:LSsim} illustrate the structure of the LMC stellar disk, seen 
face on, at various times in the SMC's orbital history (marked as red stars in the 
top left panel).  Only particles initially associated with the LMC are plotted.  The location of the SMC 
is denoted by a blue star. 
The initial state of the LMC, as a symmetric disk with flocculant spiral structure
is illustrated in the panel marked T=0.7 Gyr. As the SMC orbits about the LMC, asymmetric spiral 
structure is induced. After 4.6, 5.7, and 6.3 Gyr of evolution the SMC is roughly 25 kpc from the 
LMC, which is the approximate separation of the Clouds today. The last panel illustrates the 
final state of the system, after the SMC has been completely consumed by the LMC.

% l b r t

As seen in Figure~\ref{fig:LSsim}, a dominant one-armed spiral is induced after each pericentric approach 
of the SMC, but it is strongest after the SMC passes through the disk itself, particularly at
pericenter approaches $<$ 10 kpc (e.g. T=6.3).  The structure in the panel marked T=6.3 Gyr is most 
reminiscent to the observed structure in the LMC. 
 Multiple spiral arms are formed in the region marked B and a 
 sub-dominant arm is seen in the south, analogous to that observed (referred to as the B1 and B3
regions in deVF74). The northern-most structures reach as far as 15 kpc from the LMC center
(Region A), like the arc seen by \citet{Mac15}. 
The mass resolution of the simulation is 2500$M_\odot$/particle. This is insufficient to create 
detailed star count maps to compare
against the \citet{Mac15} observations, but the stellar density in the outskirts drops sharply
by roughly an order of magnitude 
from 10 to 20 kpc, also as observed (Region B).  
This simulation illustrates that 
Milky Way tides are not necessary to form low density stellar arcs in the outskirts of the LMC
that are similar to those observed.

The strength, asymmetry and number of observed spiral arms can thus be used to constrain the 
impact parameter of the most recent LMC/SMC interaction. From this study, 
a direct collision ($b <$10 kpc) appears necessary to reproduce the observed asymmetry of the outer 
spiral morphology of the LMC.  The structure induced from a wide separation passage (T =4.6 Gyr) are
more symmetric, reminiscent of the simulations presented in \citet{Mac15} of the impact of global Milky Way tides.

\subsection{The LMC and SMC about the Milky Way} 
\label{subsec:LSMW}

% l b r t
\begin{figure*}
\begin{center}
\mbox{\includegraphics[width=4in, angle=-90 ]{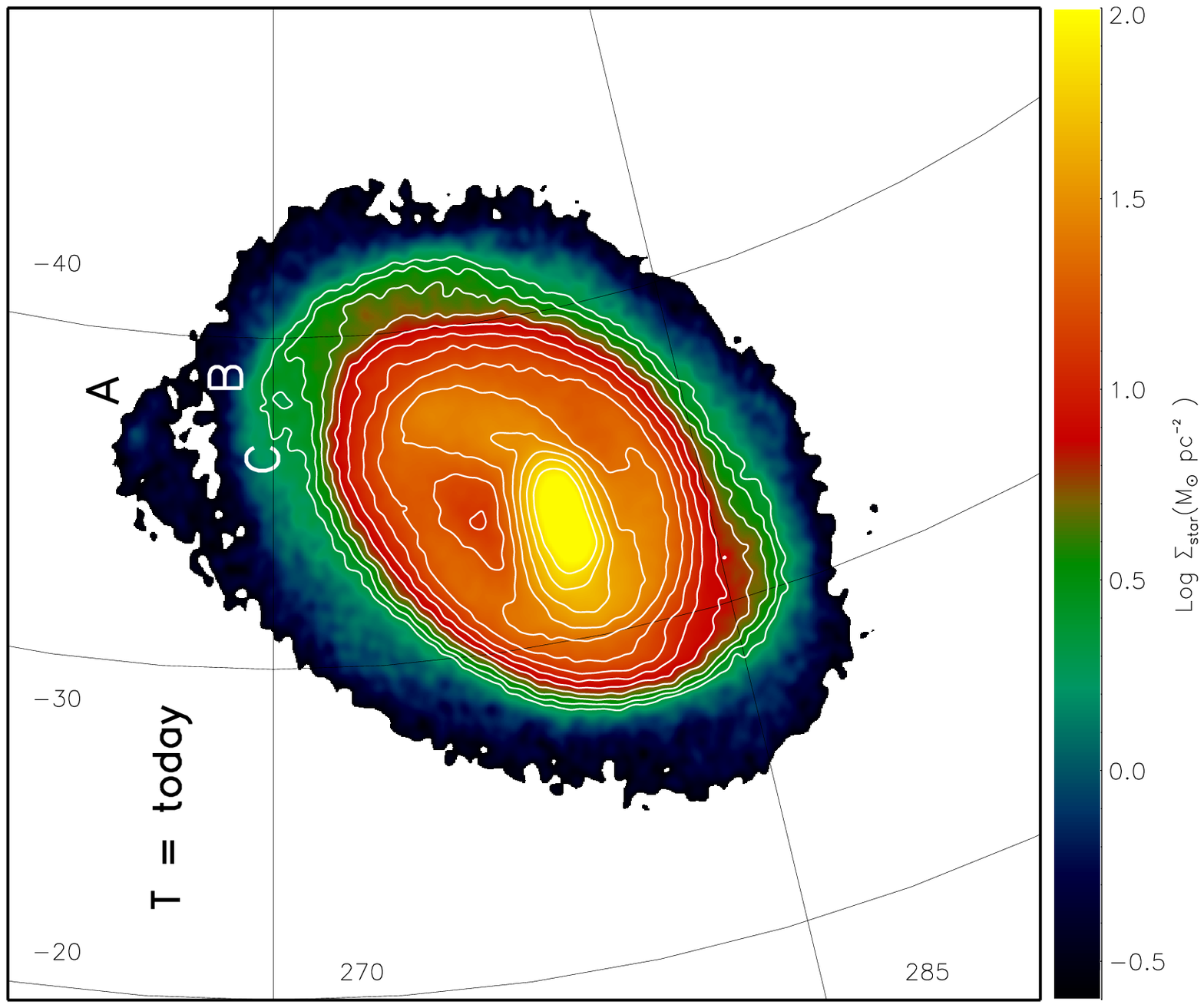}
\includegraphics[width=3.4in, angle = -90, clip=true, trim = 0 0 0.99in 0]{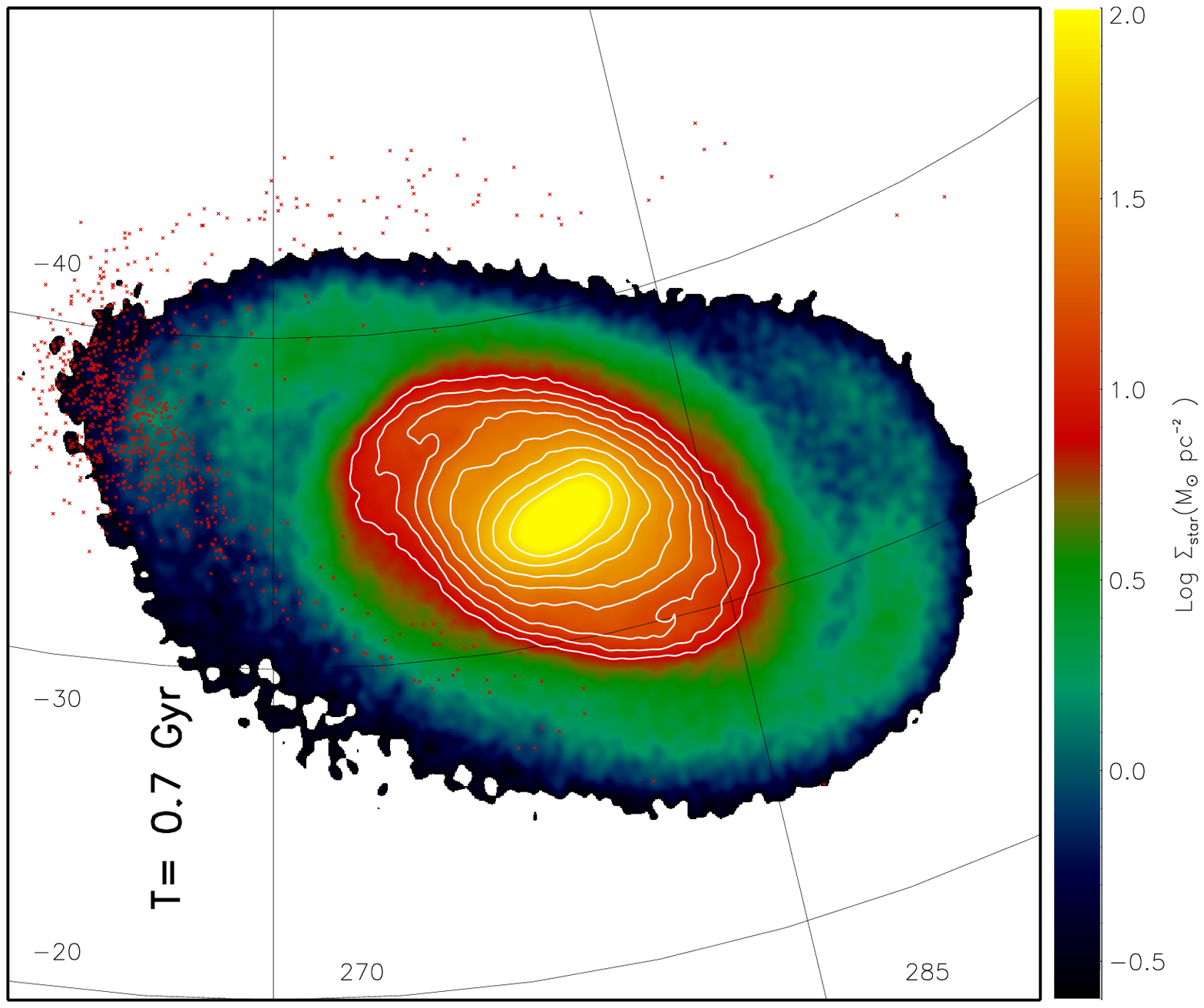}}
 \end{center}
 \caption{\label{fig:LMW} Left: Stellar density of the simulated LMC disk from B12 in Galactic coordinates
 at the present time (1 Gyr after the LMC crossed R200 of the Milky Way). 
 White contours highlight the inner disk structure, which consists of a one-armed spiral
 and off-center stellar bar in the higher density regions (red) and arcs in the lower density region C, 
 analogous to those revealed by our deep imaging 
 campaign (Figure~\ref{fig:LMC}). The northern-most stellar arc (Region A) is similar to the low-density structure identified by 
 \citet{Mac15} and is located $\sim15$ degrees from the LMC center.  The stellar density 
 drops sharply in Region B.  Right: The simulated LMC disk
  0.7 Gyr ago, well before the pericentric approach of the LMC to the MW. The disk has been 
 placed at the current location of the LMC to facilitate a line of sight comparison with the present day disk. As such, 
 lines of Galactic latitude are only drawn for reference. 
 Red dots illustrate the location of stellar particles that ultimately end up in arc A. These red particles
 are located in arms of the outer disk that were perturbed by the SMC at distances where 
 Milky Way tides are negligible.    }
 \end{figure*}

In the previous section we illustrated that close interactions between the LMC and SMC ($b< 10$kpc)
 are sufficient to create asymmetric spiral arms and arcs in the outskirts of the LMC disk that are similar 
 to those observed.  Here we include the Milky Way to this picture to assess the degree to which
 the tidal field of the Galaxy is affecting this picture. 
 
 We turn to simulations of the Clouds and Milky Way that reproduce both the internal structure of the LMC
 and the major components of the extended Magellanic System (i.e., Bridge and Stream):  
Model 2 of B12.   In this model the Clouds have just passed their first pericentric approach 
to a $1.5 \times 10^{12}$ M$_{\odot} $   Milky Way \citep[a distance of $\sim$49 kpc][]{B07}
 and the SMC has recently collided with the LMC $\sim 0.1$ Gyr ago to form the
Magellanic Bridge (impact parameter $b < 5$ kpc).   As illustrated in Figure 5 and 6 of \cite{B13}, there is stellar debris from 
the SMC surrounding the LMC disk \citep[see also][]{O11} and also strong perturbations to both the outskirts and inner regions
 of the LMC disk in this model. 
 
In the left hand panel of Figure~\ref{fig:LMW} we plot the stellar density of particles originally associated with the LMC at
the present time in Galactic coordinates (1 Gyr after the LMC crosses R200 of the Milky Way). 
Debris from the SMC is predicted to be more diffusely distributed behind and about the LMC disk \citep{B13} and 
does not contribute to the prominent stellar arcs seen in the outskirts of the simulated LMC disk.  
White contours highlight structures in the inner disk, including a one-armed spiral and 
geometrically off-center bar - these structures do not form without a low impact parameter collision between the 
Clouds \citep[B12]{B03}.    The green region (5-8 degrees north of the center of the bar, marked C) is analogous to the 
northern structures visible in our imaging campaign (Figure~\ref{fig:LMC} and ~\ref{fig:Panorama}). 
The density of the disk drops sharply outside a radius of 10 degrees (Region B). 
The Northern-most arc structure 
(marked A, located $\sim$ 15 degrees from the bar) is an order of magnitude less dense than region B 
 and may be similar to the structure observed by 
\citet{Mac15}.  There is no Southern counterpart to this structure.  These structures are analogous to those 
described in Figure~\ref{fig:Mac}, where Clouds interact in isolation (no Milky Way).

Here, tidal forces from both the SMC and Milky Way are acting on the LMC disk at this point in the simulation. 
Milky Way tides must play a role in shaping the stellar debris at the largest radii, however that does 
not imply that tides from the Galaxy originally formed the structures. 
 To determine the origin of the northern-most stellar arc (A) in the left-hand
panel of Figure~\ref{fig:LMW}, we tag particles associated with this structure at the present time and trace them 
back to an earlier point in time.  

In the right hand panel of Figure~\ref{fig:LMW} we plot the stellar density of the 
LMC disk $\sim$0.7 Gyr ago ($\sim$0.3 Gyr after the LMC crossed R200 of the Milky Way).  At this point Milky 
Way tides are minimal as the LMC is located $\sim$170 kpc from the Galactic center. For the sake of comparison, 
the figure illustrates how the LMC disk would look like from the same distance and viewing perspective as the 
LMC today (left panel).   Red dots mark the location of stellar particles associated with arc A from the 
left panel.    The outskirts of the LMC disk are highly disturbed by SMC tides (recall, Milky Way tides are negligible). 
Indeed the simulated disk asymmetries are similar to those in Figure~\ref{fig:LSsim}. 

Given the much smaller mass of the SMC, it's tidal field is a localized effect, versus the global perturbations induced 
by the Milky Way, thus giving rise to the pronounced asymmetry in the resulting tidal structures.  
While the final shape of the structure may be influenced by Milky Way tides after the recent pericentric approach of the LMC, 
the stellar arc originates from LMC stars in the outskirts of the disk that are initially tidally disturbed by the SMC during 
previous passages, implying the existence of a long-lived LMC-SMC binary system.

While writing this paper it became clear that the LMC disk model in the B12 was not inclined correctly 
with regards to our line of sight.  The inclination has been corrected for the line-of-sight images shown in 
Figure~\ref{fig:LMW}; however it should be noted that as a result the SMC is not in the correct location on the plane of sky. 
As such, the exact orientation of the stellar arcs in the northern regions should not be interpreted as precise predictions.  
Also, our understanding of the 3D velocities of the LMC have changed since the B12 simulations were created \citep{K13}.  
But a first infall scenario
 is still the most plausible scenario for Milky Way mass models with virial masses $<1.5 \times 10^{12}$ M$_{\odot}$ \citep{K13}. 
 As such, the change in velocity does not affect the general conclusions drawn in this section.

\section{Discussion: Connection to Magellanic Irregular Galaxies}
\label{sec:Discuss}
	
Many Magellanic Irregular dwarf galaxies, i.e. analogs of the LMC, 
are known to have low mass companions \citep[Pearson et al., submitted]{W04,O94}. 
Classic examples include NGC 4490/85, UGC 9562/60, NGC 3478/UGC 6016.  
Each of these dwarf pairs also possess connecting HI bridges, analogous to the 
Magellanic Bridge \citep[]{C98,N86,C01} and extended HI envelopes that are 
unlikely to arise owing to environmental effects (Pearson, Besla, submitted).
 It is thus postulated that the off-center stellar bar and highly asymmetric  
 spiral structure that characterize
the Magellanic Irregular galaxy class \citep[]{deV72} may result from 
interactions with a companion dwarf galaxy \citep[]{B12,B03,YB14}. 
Indeed many Magellanic Irregular galaxies are not in proximity to a massive host galaxy (e.g. Arp22).

The results shown in Figures~\ref{fig:LSsim} and ~\ref{fig:LMW} are consistent with
 previous theoretical studies of minor mergers
and we expect them to be generically applicable to LMC analogs in the field. 
For example, \citet[]{B03} modeled the stellar structure of barred Milky Way like galaxies
during collisions with smaller galaxies. They
similarly conclude that off-center impacts can induce
one-armed spiral structure \citep[see also similar studies by: ][]{S97,A97,WH93,LT92}.

Figure~\ref{fig:LSsim} further illustrates that extreme
impact parameters ($<$ 5 kpc) are not a necessary condition to reproduced the broad morphology
of the outer stellar disks of Magellanic Irregulars. In the specific case of the LMC, at least one close impact is likely required 
to reproduce the LMC's inner disk morphology in detail, particularly that of the stellar bar. 
B12 showed that an off-center SMC impact (b $\sim$2-5 kpc) can warp
the bar out of the LMC disk plane, consistent with observations \citep{Haschke2012}. A 10 kpc impact, as illustrated by 
the panel marked T=6.3 Gyr, is able to reproduce the outer spiral morphology, but the bar is still
coplanar with the disk.

A major argument against the B12 picture that interactions between dwarfs are responsible for
 the disturbed nature of Magellanic Irregulars 
is that many do not have obvious companions \citep[]{W04}. But, perhaps surprisingly, 
Figure~\ref{fig:LSsim} illustrates that, while many of the outer stellar structures are 
transient,  the dominant one-armed spiral persists even after the SMC is completely cannibalized at T=8.0 Gyr!
A 1:10 mass ratio merger\footnote{Note that the mass ratio
is larger at the final stages of coalescence as much of the dark matter mass of the SMC was 
placed in an extended halo, which is easily truncated by LMC tides.} is insufficient to destroy
the disk of the LMC. This result is consistent with results from studies of minor mergers on 
disk stability for more massive, gas rich systems \citep[e.g.,][]{M12,H08,B05}. 
As such, the final merged system, when signs of a companion are minimal, maintains the appearance of a
Magellanic Irregular galaxy. 

The one-armed spiral structure dissipates within 1-2 Gyr of 
coalescence. These structures are thus hallmarks of a very recent or ongoing minor merger.
Indeed, the Magellanic Irregular galaxy NGC 4449 was thought 
to be isolated until a low surface brightness companion was found in its periphery \citep[]{MD12}. 
We expect that similar deep imaging campaigns about other Magellanic Irregulars are likely to
reveal stellar signatures of tidal disturbances.  

The {\it TiNy Titans Survey} (TNT), a multi-wavelength survey of the gas and star formation properties of 
pairs of dwarf galaxies \citep{Stierwalt15},  demonstrated that star formation is enhanced in paired dwarfs over
isolated counterparts.  Our study further develops this picture, highlighting the role that dwarf-dwarf
interactions can play in the morphological evolution of these low mass systems. 
The expected frequency of such interactions can be estimated from cosmological simulations, 
and is the subject of ongoing work (Besla et al. 2016 in prep).

\section{Conclusions}

Recent deep observations of the stellar periphery of the LMC have been published using data from the 
Dark Energy Survey identifying a stellar arc or stream in the very outskirts of the LMC (r $\sim15$ degrees
from the center) \citep{Mac15}.  These authors suggest that Milky Way tides may be responsible 
for the origin of  this structure.  However, the DES footprint does not cover the southern regions of the 
LMC outer disk, where the \citep{Mac15} models predict that complementary structures should exist.   
Furthermore, these data do not examine the structure of the LMC disk from radii of 5-10 kpc which 
could influence the origin of structure in the very outskirts.

In this study we present deep, high-resolution, optical images of the Magellanic System using 
low cost robotic telescopes with wide fields of view to explore stellar substructure in the outskirts of the 
stellar disk of the LMC (r $< 10$ degrees from the center).  These high resolution data build upon the 
seminal work of deVF72, which were the first team to identify that substantial structure exists in the 
outskirts of the LMC.  Our data has confirmed the existence of stellar arcs and multiple
spiral arms in the northern periphery, with no comparable counterparts in the South.  The asymmetry 
of these structures disfavors a formation scenario driven by global Milky Way tides.  Galactic cirrus
is minimal in the North and unlikely to affect interpretation of these observations. 

We have compared these data to detailed simulations of the
LMC disk resulting from interactions with its low mass companion, the SMC, in isolation and with 
the inclusion of the Milky Way tidal field in order to assess the origin of these northern structures, including
the \citet{Mac15} stellar arc at  $\sim $15 degrees.  

We conclude that repeated close interactions with the SMC are primarily responsible for the asymmetric 
stellar structures induced in the periphery of the LMC, particularly the dominant one-armed spiral and multiple
spiral arms or arcs in the northern disk. This is clearly illustrated in the panel marked T=6.3 Gyr in 
Figure~\ref{fig:LSsim}, where the LMC is seen to be highly distorted owing to the tidal 
field of the SMC alone.
 
 While Milky Way tides likely influence the final 
distribution of structures in the outskirts, the origin of structures such as the 
distant arc identified by \citet{Mac15} is likely a relic of the past LMC-SMC interaction history. Testing 
this model would require similarly deep observations of the southern outskirts of LMC disk; comparable 
structures should exist if the origin of this structure is Milky Way tides. 
The upcoming DECam Magellanic Satellites Survey 
(PI: Bechtol) will extend the DES footprint to the south and should shed more light on this scenario. 

These new data, particularly the properties of the arcs in the North, thus provide powerful new 
constraints that will allow us to narrow down the orbital parameter space and understand in
detail the recent interaction history of the Magellanic Clouds. In particular, these data 
can help constrain the longevity of the LMC/SMC binary state.  Such comparisons of high resolution, 
deep optical data and 
 simulated stellar structure of a generic 1:10 mass ratio dwarf galaxy encounter  provides
 a strong illustration of how dwarf-dwarf interactions can 
fundamentally modify the stellar structure of dwarf galaxies like the LMC.

Outer spiral arms and arcs are found to be transient, but the dominant one-armed spiral induced 
by a collision between dwarfs would persist for 1-2 Gyr even after the secondary merges entirely 
with the primary. As such, the lack of a companion around a Magellanic Irregular does not disprove
the hypothesis that the structures defining this class of galaxy (one-armed spiral and geometrically
off-center bar) are driven by interactions with low mass companions. Instead, such stellar structures are
hallmarks of ongoing interactions or a recent merger event. Deep, wide-field observations of the 
outer peripheries of 
Magellanic Irregulars may thus harbor clues to test this theoretical picture.

\section{Acknowledgements}

We thank David Schminovich for useful conversation that have improved this manuscript. 
GB acknowledges support through HST AR grant \#12632.  Support for program \#12632 was provided 
by NASA through a grant from the Space Telescope Science Institute, which is operated by the 
Association of Universities for Research in Astronomy, Inc., under NASA contract NAS 5-26555.
 DMD and EKG acknowledge support by Sonderforschungsbereich (SFB) 881 ``The Milky Way System'' of 
 the German Research Foundation (DFG), particularly through subproject A2. 
ES acknowledges support for this work provided by NASA through Hubble
Fellowship grant HST-HF2-51367.001-A awarded by the Space Telescope
Science Institute, which is operated by the Association of
Universities for Research in Astronomy, Inc., for NASA, under contract
NAS 5-26555. The simulations in this paper were produced on the Odyssey cluster supported by 
the FAS Science Division Research Computing Group at Harvard University. 
Analysis was undertaken on the El Gato cluster at the University of Arizona, 
which is supported by the 
National Science Foundation under Grant No. 1228509


\begin{thebibliography}{}

\bibitem[Abraham \& van Dokkum (2014)]{Ab14} Abraham, R.G. \& van Dokkum, P.G. 2014, \pasp, 126, 55
\bibitem[Amorisco et al.(2015)]{AM15} Amorisco, N.C., Mart\'{i}nez-Delgado, D., \& Schedler, J. 2015 arXiv:1504.03697
\bibitem[Athanassoula et al.(1997)]{A97} Athanassoula, E., Puerari, I., Bosma, A., 1997, \mnras, 286, 284
\bibitem[Balbinot et al.(2015)]{Balbinot15} Balbinot, E., et al., 2015, \mnras, 449, 1129
\bibitem[Bechtol et al.(2015)]{DES1} Bechtol, K. et al. 2015, \apj, 807, 50
\bibitem[Belokurov \& Koposov(2016)]{BK16} Belokurov, V. \& Koposov, S. 2016, \mnras, 456, 602
\bibitem[Berentzen et al.(2003)]{B03} Berentzen, I., Athanassoula, E.; Heller, C. H.; Fricke, K. J, 2003, \mnras, 342, 343
\bibitem[Besla et al.(2007)]{B07} Besla G., Kallivayalil N., Hernquist L., Robertson B., Cox T. J., van der Marel R. P., Alcock C., 2007, \apj, 668, 949
\bibitem[Besla et al.(2010)]{B10} Besla, G., Kallivayalil, N., Hernquist, L., van der Marel, R.P., Cox, T.J., Kere\~{s}, D. 2010, \apj, 721, 97
\bibitem[Besla et al.(2012)]{B12} Besla, G., Kallivayalil, N., Hernquist, L., van der Marel, R.P., Cox, T.J., Kere\~{s}, D.  2012, \mnras, 421, 2109
\bibitem[Besla et al.(2013)]{B13} Besla, G., Hernquist, L., Loeb, A., 2013, \mnras, 428, 2342
\bibitem[Besla (2015)]{B15} Besla, G. 2015, arXiv:1511.03346
\bibitem[Bica et al.(2008)]{Bica08} Bica, E., Bonatto, C., Dutra, C. M., Santos, J. F. C., 2008, \mnras, 389, 678
\bibitem[Bournaud et al.(2005)]{B05} Bournaud, F., Jog, C. J., Combes, F., 2005, A\&A, 437, 69
\bibitem[Boylan-Kolchin et al.(2011)]{BK11} Boylan-Kolchn, M., Besla, G., \& Hernquist, L. 2011, \mnras, 414, 1560
\bibitem[Casetti-Dinescu et al.(2012)]{CD12} Casetti-Dinescu, D.I., Vieira, K., Girard, T.M., \& van Altena, W.F. 2012, \apj, 753, 123
\bibitem[Clemens et al.(1998)]{C98} Clemens M. S., Alexander P., Green D. A., 1998, \mnras, 297, 1015
\bibitem[Cox et al.(2001)]{C01} Cox, A.L, Sparke, L.S., Watson, A.M., \& van Moorsel, G. 2001, \aj, 121, 692
\bibitem[Diaz \& Bekki(2011)]{DB11} Diaz, J., \& Bekki, K. 2011, \mnras, 413, 2015
\bibitem[Drlica-Wagner et al.(2015)]{DES2} Drlica-Wagner, A., et al. 2015,  \apj, 813, 109
\bibitem[de Vaucouleurs \& Freeman(1972)]{deV72} de Vaucouleurs, G., Freeman, K. C., 1972, Vistas Astron., 14, 163
\bibitem[Duc et al.(2015)]{Duc15} Duc, P.-A., et al. 2015, \mnras, 446, 120
\bibitem[G\'{o}mez, F.  et al.(2015)]{G15} G\'{o}mez, F., Besla, G., Carpintero, D.D., Villalobos, \'{A}lvaro, O'Shea, B.W., \& Bell, E.F. 2015, \apj, 802, 128
\bibitem[Graff et al.(2000)]{Graff2000} Graff, D.S., Gould, A.P., Suntzeff, N.B., Schommer, R.A., \& Hardy, E. 2000, \apj, 540, 211
\bibitem[Harris(2007)]{H07} Harris, J., 2007, \apj, 658, 345
\bibitem[Haschke et al. (2012)]{Haschke2012} Haschke, R., Grebel, E.K., \& Duffau, S. 2012,  \aj, 144, 106
\bibitem[Hindman et al.(1963)]{H63} Hindman, J.V., et al., 1963, Australian Journal of Physics, 16, 570
\bibitem[Hopkins et al.(2008)]{H08} Hopkins, P. F., Hernquist, L., Cox, T. J., Younger, J. D., Besla, G., 2008, \apj, 688, 757
\bibitem[Kallivayalil et al.(2013)]{K13} Kallivayalil, N., van der Marel, R. P., Besla, G., Anderson, J., Alcock, C., 2013, \apj, 764, 161
\bibitem[Kapakos \& Hatzidimitriou(2012)]{KH12} Kapakos, E., Hatzidimitriou, D., 2012, \mnras, 426, 2063
\bibitem[Kerr(1957)]{K57} Kerr F. J., 1957, \aj, 62, 93
\bibitem[Kontizas et al.(1990)]{K90} Kontizas, M., Morgan, D. H., Hatzidimitriou, D., Kontizas, E., 1990, A\&AS, 84, 527 
\bibitem[Koerwer (2009)]{K09} Koerwer, J.F., 2009, \aj, 138, 1
\bibitem[Koposov et al.(2015)]{K15} Koposov, S.E., Belokurov, V., Torrealba, G., Evans, N. W. 2015, \apj, 805, 130
\bibitem[Kunkel et al.(1997)]{Kunkel97} Kunkel, W.E., Demers, S., Irwin, M.J., \& Albert, L. 1997, \apj, 488, L129
\bibitem[Lah et al. (2005)]{Lah05} Lah, P., Kiss, L.L., Bedding, T.R. 2005, \mnras, 359, L42
\bibitem[Lynds \& Toomre(1972)]{LT92} Lynds R., Toomre A., 1976, \apj, 209, 382
\bibitem[Mackey et al.(2015)]{Mac15} Mackey, D., Koposove, S.E., Erkal, D., Belokurov, V., Da Costa, G.S., \& G\'{o}mez, F.A. 2015 arXiv 15508.1356
\bibitem[Martin et al.(2015)]{M15} Martin, N.F., et al. 2015, \apj, 804, L5
\bibitem[Mart\'{i}nez-Delgado et al.(2012)]{MD12} Mart\'{i}nez-Delgado, D., et al. 2012, \apj, 748, L24
\bibitem[Mart\'{i}nez-Delgado et al.(2010)]{MD10} Mart\'{i}nez-Delgado, D., et al. 2010, \aj, 140, 962
\bibitem[Mart\'{i}nez-Delgado et al.(2008)]{MD08} Mart\'{i}nez-Delgado, D., Pe\~{n}arrubia, J., Gabany, R. J., et al., 2008, ApJ, 689, 184
\bibitem[Mathewson et al.(1974)]{M74} Mathewson, D.S., Cleary, M.N., \&  Murray, J.D. 1974, \apj, 190, 291
\bibitem[McMonigal et al.(2014)]{Mc14} McMonigal, B., Bate, N.F., Lewise, G.F., Irwin, M.J., Battaglia, G., Ibata, R.A., Martin, N.F., 
McConnachie, A.W., Guglielmo, M., \& Conn, A.R. 2014, \mnras, 444, 3139
\bibitem[Meschin, et al.(2014)]{Meschin14} Meschin, I., Gallart, C., Aparicio, A., Hidalgo, S.L., Monelli, M., Stetson, P.B., \& Carrera, R. 2014, \mnras, 438, 1067
\bibitem[Mihos et al.(2005)]{Mihos05} Mihos, C. J., Harding, P., Feldmeier, J., \& Morrison, H., 2005, \apj, 631, 41
\bibitem[Moretti et al.(2014)]{Moretti14} Moretti, M.I., Clementini, G., Muraveva, T., Ripepi, V., marquette, J.B., Cioni, M.-R. L., Marconi, M., Girardi, L., Rubele, S., Tisserand, P., 
de Grijs, R., Groenewegen, M.A.T., Guandalini, R., Ivanov, V.D., \& van Loon, J. Th. 2014, \mnras, 437, 2702
\bibitem[Moster et al.(2012)]{M12} 	Moster, B. P., Macc\'{o}, A. V., Somerville, R. S., Naab, T., Cox, T. J., 2012, \mnras, 423, 2045
\bibitem[Moster et al.(2013)]{M13} Moster, B.P., Naab, T., \& White, D.M. 2013, \mnras, 428, 3121
\bibitem[Mu\~{n}oz et al.(2006)]{Mun06} Mu\~{n}oz, R.R., et al. 2006, \apj, 649, 201
\bibitem[Nidever et al.(2010)]{N10} Nidever, D.L., Majewsky, S.R., Burton, W.B., and Nigra, L. 2010,\apj, 723, 1618
\bibitem[Nidever et al.(2011)]{N11} Nidever, D. L., Majewski, S.R.  Mu\~{n}oz, R.R., Beaton, R.L., Patterson, R.J., and Kunkel, W.E., 2011, \apj, 733L, 10
\bibitem[Nidever et al.(2013)]{N13} Nidever, D.L., Monachesi, A., Bell, E. F., Majewski, S. R., Mu\~{n}oz, R. R., Beaton, R. L., 2013, \apj, 779, 145
\bibitem[Nikolaev et al.(2004)]{N04} Nikolaev, S., Drake, A.J., Keller, S.C., Cook, K.H., Dalal, N., Griest, K., Welch, D.L., Kanbur, S.M., 2004, \apj, 601, 260
\bibitem[Noreau \& Kronberg(1986)]{N86} Noreau, L., Kronberg, P. P., 1986, \aj, 92, 1048
\bibitem[Odewahn(1994)]{O94} Odewahn S. C., 1994, \aj, 107, 1320
\bibitem[Olsen et al.(2011)]{O11} Olsen, K.A.G., Zaritsky, D., Blum, R.D., Boyer, M.L., Gordon, K.D., 2011, \apj, 737, 29
\bibitem[Olsen \& Salyk(2002)]{OS02} Olsen, K.A.G., \& Salyk, C. 2002, \apj, 656, L61
\bibitem[Pe\~{n}arrubia et al. (2016)]{Pen16} Pe\~{n}arrubia, J., G\'{o}mez, F., Besla, G., Erkal, D., \& Ma, Y.-Z. 2016, \mnras, 456, 54
\bibitem[Putman et al.(2003)]{P03} Putman M. E., Staveley-Smith L., Freeman K. C., Gibson B. K., Barnes, D. G., 2003, \apj, 586, 170
\bibitem[Rubele et al.(2015)]{Rubele15} Rubele, S. et al. 2015, \mnras, 449, 639
\bibitem[Saha et al.(2010)]{S10} Saha, A., et al., 2010, \aj, 140, 1719
\bibitem[Salem et al.(2015)]{Salem15} Salem, M., Besla, G., Bryan, G., Putman, M., van der Marel, R.P., \& Tonnesen, S. 2015, \apj, 815, 77
\bibitem[Schlegel et al.(1998)]{S98} Schlegel, D.J., Finkbeiner, D.P., \& Davis, M. 1998, \apj, 500, 525
\bibitem[Springel(2005)]{S05} Springel, V., 2005, \mnras, 364, 1105
\bibitem[Stierwalt et al.(2015)]{Stierwalt15} Stierwalt, S., Besla, G., Patton, D., Johnson, K., Kallivayalil, N., Putman, M., Privon, G., Ross, G. 2015, \apj, 805, 2
\bibitem[Struck(1997)]{S97} Struck C., 1997, ApJS, 113, 269
\bibitem[Subramanian(2003)]{Sub03} Subramaniam, A. 2003, \apj, 598, L19
\bibitem[Udalski et al.(2015)]{U15} Udalski, A., Szyma\'{n}ski, MK., \& Szyma\'{n}ski, G. 2015, Acta Astronomica, 65, 1, 1
\bibitem[van der Marel(2001)]{vdM01} van der Marel, R.P., 2001, \aj, 122, 1827
\bibitem[van der Marel \& Cioni(2001)]{vdMC01} van der Marel, R.P., \& Cioni, M.-R. L., 2001, \aj 122, 1807
\bibitem[van Dokkum et al. (2014)]{vDok14} van Dokkum, P.G., Abraham, R., \& Merritt, A. 2014, \apj, 782, 24
\bibitem[van Dokkum et al. (2015)]{vDok15} van Dokkum, P.G., Abraham, R., Merritt, A., Zhang, J., Geha, M., \& Conroy, C. 2015, \apj, 798, 45
\bibitem[Weil \& Hernquist(1993)]{WH93} Weil M. L., Hernquist L., 1993, \apj, 405, 142
\bibitem[Wilcots \& Prescott(2004)]{W04} Wilcots E. M., Prescott M. K. M., 2004, \aj, 127, 1900
\bibitem[Yozin \& Bekki(2014)]{YB14}Yozin, C. \& Bekki, K. 2014, \mnras, 439, 1948

\end{thebibliography}
\end{document}